\documentclass{aa}  
\usepackage[varg]{txfonts}

\usepackage{natbib}
\usepackage{amsmath}
\usepackage{float}
\bibpunct{(}{)}{;}{a}{}{,} 
\usepackage{graphicx}
\usepackage{txfonts}
\usepackage{xcolor}
\usepackage[euler]{textgreek}

\usepackage[options]{hyperref}

\begin{document}

\title{Do all gaps in protoplanetary discs host planets?}

             \author{Anastasia Tzouvanou \thanks{anastasiatzouvanou@gmail.com}
          \inst{1}
          \and Bertram Bitsch
          \inst{1}
          \and Gabriele Pichierri
          \inst{1,2}
            }

   \institute{Max-Planck-Institut für Astronomie, Königstuhl 17, 69117 Heidelberg, Germany
              \\
         \and
         GPS Division, Caltech, Pasadena, California
             \\
             }

   \date{Received; accepted }

% \abstract{}{}{}{}{} 
% 5 {} token are mandatory
 
  \abstract{Following the assumption that the disc substructures observed in protoplanetary discs originate from the interaction between the disc and the forming planets embedded therein, we aim to test if these putative planets could represent the progenitors of the currently observed giant exoplanets. 
  We performed N-body simulations assuming initially three, four, five or seven planets. Our model includes pebble and gas accretion, migration, damping of eccentricities and inclinations, disc-planet interaction and disc evolution. We locate the planets in the positions where the gaps in protoplanetary discs have been observed and we evolve the systems for 100Myr including a few Myr of gas disc evolution, while also testing three values of $\alpha$ viscosity.
  For planetary systems with initially three and four planets we find that most of the growing planets lie beyond the RV detection limit of 5AU and only a small fraction of them migrate into the inner region. We also find that these systems have too low final eccentricities to be in agreement with the observed giant planet population.  Systems initially consisting of five or seven planets become unstable after $\approx$40Kyr of integration time. This clearly shows that not every gap can host a planet. The general outcome of our simulations - too low eccentricities - is independent of the disc's viscosity and surface density. Further observations could either confirm the existence of an undetected population of wide-orbit giants or exclude the presence of such undetected population to constrain how many planets hide in gaps even further.}


\maketitle

\section{Introduction}
High-resolution observations of dust in protoplanetary discs by the Atacama Large Millimeter/submillimeter Array (ALMA) have revealed complex features in form of gaps and rings, arcs and spirals, commonly referred to as protoplanetary disc substructures \citep{bae2022structured}. Since then, many studies have focused on these disc substructures and their relation to disc evolution and planet formation. Theoretical explanations that have been proposed to explain these features include, among others, pebble growth near condensation zones \citep{zhang2015evidence}, zonal flows \citep{flock2015gaps}, self-induced dust traps \citep{gonzalez2017self}, gravitational instabilities \citep{takahashi2016origin} or the interaction between the disc and the embedded growing planets (e.g. \citealt{pinilla2012ring,long2018gaps}).

In some cases, there is additional evidence, derived from gas kinematics, for the presence of growing planets in the observed gaps of protoplanetary discs \citep{muller2018orbital,keppler2018discovery,benisty2021circumplanetary}. By measuring the rotation curves of CO emission, \citep{teague2018kinematical} showed that the perturbations of the gas surface density in the HD 163296 disc are likely caused by the presence of two Jupiter-mass planets. Similar work by \citep{pinte2018kinematic} also found that a two Jupiter mass planet in the HD 163296 disc located at 260AU can reproduce the observed CO emission curve.
    
For these reasons, many authors have investigated the possibility that the observed substructures in discs are generally caused by growing planets, which would allow to derive indirect information on the early stages of the planet-formation process. Note that the DSHARP project has detected disc substructures across the range of 5-150AU \citep{huang2018disk}, while the inner regions of protoplanetary discs, that is at distances where fully-formed exoplanets are more readily detected, remain difficult to resolve \citep{andrews2016ringed}. Indeed, there are strong observational biases to keep in mind, with limitations in terms of orbital separations as well as disc and stellar masses \citep{bae2022structured}. Still, if the observed rings/gaps are indeed caused by the presence of planets, their existence implies that they should continue to grow and migrate, potentially into regions of the parameter space that are accessible by radial velocity (RV) surveys. These putative planets would then constitute the progenitors of the RV exoplanet sample observed today. 

Several studies have been performed in order to investigate the origin of gaps/rings in protoplanetary discs by examining planet-disc and planet-planet interactions and relate current models with exoplanet observations. \citep{lodato2019newborn} studied 48 gaps in the context of planet-disc interactions. Assuming the same fixed disc properties and disc lifetime for all the investigated gaps, they evolved and migrated the planets for 3-5Myr. After 3-5Myr they found that the final distribution of planets in a mass-semi major axis plot is consistent with the observed exoplanet distribution, in particular for cold Jupiters. A similar study, again assuming one planet per disc, was carried out by \citep{ndugu2019observed}. They  studied the evolution of planets above pebble isolation mass for a few Myr and they compared their resulting population with observations. In contrast to \citep{lodato2019newborn}, \citep{ndugu2019observed} find that their planet population is not in agreement with giant planet observations. The main differences arise from the different viscosity values used to model the disc: the low viscosity of the study by \citep{ndugu2019observed} prevents the planets from migrating into orbital separations where they could be observed via radial velocity, while the higher viscosity in the study by \citep{lodato2019newborn} results in efficient inward migration into locations where the giants are observed today.
    
The first work also including the planet-planet interaction was done by \citep{muller2022emerging}. Instead of covering a range of observed disc substructures in a statistical sense but treated individually, they focused on the HD 163296 disc, where they assumed the presence of three interacting planets in observed gaps at 48, 83 and 137 AU; they evolved their systems for a total of 100Myr, including a few Myr lifetime of the gas disc. Their findings regarding the eccentricity distribution showed that in that system the interaction between planets did not result in large eccentricities to match the observed giant planet distribution. They also note that higher viscosity values are more favorable in order to reproduce a planet population that is consistent with current observations, in line with the single body simulations \citep{ndugu2019observed,lodato2019newborn}.

Our work aims to also include planet-planet interactions while keeping a more general approach without focusing on a specific target. Motivated by the gaps/rings of the observed protoplanetary discs, we run N-body simulations with 3,4,5 and 7 planets. The initial locations and masses of the planets are informed by the positions of the gaps in the observed discs  and the local pebble isolation mass (the mass at which a planet is able to create a sufficiently strong pressure gradient to trap dust particles outside its orbit) and we study how these systems would continue to grow, migrate and gravitationally interact for a total of 100Myr (including a few Myr of gas disc evolution). Our goal is to then compare our resulting systems to the observed exoplanet distributions to see if these forming planets could indeed represent the progenitors of the observed exoplanets.

The paper is organised as follows. In Section \ref{methods} we describe our N-body code FLINSTONE and discuss the different parameters we used for our disc model as well as the initial conditions of our simulations \ref{initial_conditions}. In Section \ref{results} we present the results of our simulated systems with 3, 4, 5 and 7 planets in \ref{res:3pl}, \ref{res:4pl} and \ref{res:5-7pl} respectively. In Section \ref{discussion} we discuss our findings for the different planetary systems and we finally summarize in Section \ref{summary} the main outcomes. 

\section{Method} \label{methods}
For our simulations we used the N-body code FLINTSTONE, which includes pebble and gas accretion, gas-driven planetary migration and eccentricity/inclination damping and disc evolution (\citealt{bitsch2019formation,izidoro2017breaking}). The integrations were performed by the original N-body integrator that is based on the Mercury hybrid symplectic integrator \citep{chambers1999hybrid}. Collisions between planets were considered as inelastic merging events. In the following we describe the different parts of the model and our set-up.

\subsection{Disc model}
The gas surface density profile $Σ_{\mathrm{g}}$ of the disc is radius and time dependent and can be described as 
\begin{equation} \label{eq:surf_den}
\begin{aligned}
\Sigma_{\mathrm{g}}(r,t) = & \Sigma_{0} \cdot \Big(\dfrac{r}{r_0} \Big ) ^{-1} \exp \Big [-\Big (\dfrac{r}{r_{\mathrm{c}}} \Big ) \Big] \exp \Big[- \Big (\dfrac{t}{\tau_{\mathrm{visc}}}\Big ) \Big ]
 \\
 \\
 &   \begin{cases}
    1,   \hspace{2.7cm} t < t_{\mathrm{diss,ph.ev.}}\\
    \exp\left[ - \left( \frac{t - t_{\mathrm{diss,ph.ev.}}}{\tau_{\mathrm{diss,ph.ev.}}}\right) \right], \hspace{0.4cm}
    t > t_{\mathrm{diss,ph.ev.}} 
    \end{cases}
\end{aligned}
\end{equation}
\newline
\newline
Here $\Sigma_{0}$ is the gas surface density at $r=1$AU. The surface density profile starts to decrease exponentially with radius at the characteristic value of $r_{\mathrm{c}}$=200AU, giving a total disc mass of 0.074$M_{\odot}$.
Moreover, the surface density also decreases exponentially on a timescale of $\tau_{\mathrm{visc}}$ = 3Myrs.  To take the dissipation process during the final stages of the disc lifetime into account we assumed that the surface density profile decays exponentially with a timescale of $\tau_{\mathrm{diss,ph.ev.}}$=35Kyr from $t_{\mathrm{diss,ph.ev.}} = t_{\mathrm{disc}} - 100$Kyrs until the end of the lifetime of the disc. We summarized all the disc parameters in Table \ref{table:disc}

For the temperature profile of the disc, under the assumption of a locally isothermal state, we chose power law profile 

\begin{equation} \label{eq:temp}
    T(r) = T_{0} \cdot \left(\dfrac{r}{r_0} \right) ^{- \beta_{\mathrm{T}}},
\end{equation}
where $T_{0}=$ 150K and  $\beta_{\mathrm{T}}$ is a power law index. We chose $\beta_{\mathrm{T}}=3/7$ to achieve a flaring disc structure in the outer regions, as expected in discs dominated by stellar irradiation \citep{chiang1997spectral,dullemond2004effect,bitsch2013stellar,bitsch2015structure}:
\begin{equation} \label{eq:aspect}
    h(r) = \frac{H}{r} \propto \left( \frac{r}{r_0} \right)^{2/7}
\end{equation}

\textbf{Using the above parameters, we obtain a disc that is Toomre stable with Q parameter $ Q = \frac{c_{s} \kappa}{\pi G \Sigma_{g}} >10$ for all the radii. }

\begin{table}
\caption{Disc parameters.} 
\label{table:disc} 
\centering 
\begin{tabular}{c c } 
\hline\hline 
Parameter & Value   \\ 
\hline 
$\Sigma_{0} [g/cm^{2}]$ & $566.67$ \\
$M_{\mathrm{gas}} [M_{\odot}]$ & $0.074$ \\
$M_{\star} [M_{\odot}]$ & $1$ \\
$r_{0}$ [AU]& 1 \\
$r_{\mathrm{c}}$ [AU] & 200 \\
$r_{\mathrm{in}}$ [AU] & 0.1 \\
$r_{\mathrm{out}}$ [AU] & 500\\
$r_{\mathrm{eject}}$ [AU] & 300 \\ 
$t_{\mathrm{disc}}$ &  3-10 Myr \\
$t_{\mathrm{diss,ph.ev.}}$ & $t_{\mathrm{disc}} - 100$Kyr \\
$\tau_{\mathrm{visc}}$ & 3Myr \\
$\tau_{\mathrm{diss,ph.ev.}}$ & 1.2Kyr  \\
$\alpha$ & $10^{-3}, 3.16 \cdot 10^{-4}, 10^{-4}$ \\ 

\hline 
\end{tabular}
\end{table}

\subsection{Gas accretion}
We are interested in studying the evolution of planets that can open gaps in the pebble distribution in the disc, as we observe with ALMA. These planets need to be at least of the order of the pebble isolation mass \citep{lambrechts2014separating,bitsch2018pebble}, the mass at which the planet opens a partial gap in the disc resulting in a pressure bump exterior to its orbit, where the pebbles accumulate. We use here the pebble isolation mass prescription from \citep{bitsch2018pebble} :  
\begin{equation} \label{eq:Miso}
    M_{\mathrm{iso}} = 25f_{\mathrm{fit}}M_{\mathrm{Earth}}
\end{equation}
\vspace{0.4cm}
with 
\begin{equation} \label{eq:f_fit}
    f_{\mathrm{fit}} = \left(\frac{h}{0.05} \right)^{3} \left[0.34 \left(\frac{\mathrm{log} \big (0.001\big)}{\mathrm{log \alpha}} \right)^{4} + 0.66 \right] \left( 1 - \frac{\frac{\partial ln P}{\partial ln r}+2.5}{6}\right)
\end{equation}
where $\alpha$ is the viscosity parameter. We used the $\alpha$-disc model of \citep{shakura1973black} for which the $\alpha$ parameter is related to the turbulent viscosity as $v = \alpha H^2 \Omega_{\mathrm{K}}$. $\frac{\partial ln P}{\partial ln r}$ is the radial pressure gradient.\\
Once planets have reached the pebble isolation mass they cannot accrete pebbles anymore because the pressure bump that the planets create outside of their orbit stops further pebbles from drifting inward; thus, they can start to accrete gas into their envelope, because the heating from infalling pebbles stops. The first stage of gas accretion is regulated by the thermodynamics of the envelope, which must cool and contract in order for the planet to accrete more gas. For the contraction rate of the planetary envelope we use the same formula as in \citep{bitsch2015growth} 
\begin{equation} \label{eq:M_dotgas}
\begin{aligned}
    \dot{M}_{\mathrm{gas}} & = 0.000175  f^{-2} \left( \frac{\kappa_{\mathrm{env}}}{1 \rm{cm^{2} g^{-1}}} \right)^{-1 } \left( \frac{\rho_{\mathrm{c}}}{5.5 \rm{cm^{-3} g}} \right)^{-1/6} \left( \frac{M_{\mathrm{c}}}{M_{\mathrm{E}}} \right)^{11/3}
    \\ & \times \left( \frac{M_{\mathrm{env}}}{M_{\mathrm{E}}} \right)^{-1} \left( \frac{\mathrm{T}}{81 K} \right)^{-0.5} \frac{\rm{M_{E}}}{\rm{Myr}}
    \end{aligned}
\end{equation}
where $f$ is a factor to change the accretion rate and is set to $f = 0.2$ as in \citep{piso2014minimum} while for the opacity in the planetary envelope $\mathrm{\kappa_{env}}$ we used the same value as in \citep{bitsch2019formation} $\mathrm{\kappa_{env} = 0.05 cm^{2} g^{-1}}$. The core density is $\mathrm{\rho_{c} = 5.5 g cm^{-3}}$. If $M_{\mathrm{core}} \simeq M_{\mathrm{env}}$ the gas accretion phase regulated by cooling and contracting in hydrostatic equilibrium ends and rapid runaway gas accretion starts ($M_{\mathrm{core}} < M_{\mathrm{env}}$).
For rapid gas accretion we follow \citep{machida2010gas}, who calculated the gas accretion rate onto a planetary core using 3D hydrodynamical simulations. They found that the effective gas accretion rate is given by $ \dot{M}_{\mathrm{gas}} $ = min $\{$$\dot{M}_{\mathrm{gas,low}}, \dot{M}_{\mathrm{gas,high}}$$\}$ where the two different gas accretion limits are 
\begin{equation} \label{eq:M_dotgas}
 \begin{aligned}
    &   \dot{M}_{\mathrm{gas,low}}   =  0.83 \Omega_{\mathrm{K}} \Sigma_{\mathrm{g}} H^{2} \left(\frac{r_{\mathrm{H}}}{H} \right)^{\frac{9}{2}}\\
   & \dot{M}_{\mathrm{gas,high}} =  0.14 \Omega_{\mathrm{K}} \Sigma_{\mathrm{g}} H^{2}
\end{aligned}
\end{equation}
where $r_{\mathrm{H}}$ is the Hill radius of the planet $r_{\mathrm{H}} = r \cdot \Big ( {m_{\mathrm{pl}}}/{3M_{\star}}\Big )^{1/3}$, $Ω_{\mathrm{K}}$ is the Keplerian frequency and $Σ_{\mathrm{g}}$ is the surface density. Because the gas accretion rate of a planet depends on what the disc can provide, \textbf{in our implementation of gas accretion} we follow \citep{d2008evolution}, where they found that the gas accretion is limited to 80$\%$ of the accretion rate of the disc, which is 
\begin{equation} \label{eq:M_dotdisc}
    \dot{M}_{\mathrm{disc}} = 3\pi \alpha h^{2}r^{2}\Omega_{\mathrm{K}}\Sigma_{\mathrm{g}} \ .
\end{equation}

\subsection{Orbital migration}

Planets interact with their surrounding disc material by exchanging energy and angular momentum. The exchange of angular momentum can be explained by the wave torques in the disc \citep{goldreich1980disk}. Low-mass planets (planets around a few Earth masses), migrate in the type-I regime \citep{paardekooper2011torque} where the total torque is given as
\begin{equation} \label{eq:gamma}
   \Gamma = \Gamma_{\mathrm{L}}\Delta_{\mathrm{L}} + 
   \Gamma_{\mathrm{C}}\Delta_{\mathrm{C}}
\end{equation}
where $\Gamma_{\mathrm{L}}$ is the Lindblad torque and  $\Gamma_{\mathrm{C}}$ is the corotation torque while $\Delta_{\mathrm{L}}$ and $\Delta_{\mathrm{C}}$ are the rescaling functions of the Lindbland torque and corotation torque respectively \citep{izidoro2017breaking} to account the reduction of the total torque according to the orbital eccentricity and inclination of the planet. The nominal torques for zero eccentricity are calculated according to \citep{paardekooper2011torque}. \textbf{This migration prescription is consistent with 3D simulations \citep{bitsch2011evolution}. The same set up was used in \citep{bitsch2010orbital}, who studied the eccentricity evolution of
low mass planets and find that the damping rates in this case are consistent with the isothermal case.}\\

 When the planets are more massive, they open deep gaps in the protoplanetary discs, and they start to migrate in the type-II regime.  This happens when \citep{crida2006width}

\begin{equation} \label{eq:typeII_cond}
    \mathcal{P}= \frac{3}{4}\frac{H}{r_{\mathrm{H}}} + \frac{50}{q \mathcal{R}} \leq 1 \ .
\end{equation}
Here $q$ is the star to planet mass ratio $m_\mathrm{pl}/M_\star$ and $\mathcal{R}$ is the Reynolds number $r^{2}Ω_{\mathrm{K}}/v$. For a smooth transition from type-I to type-II migration we follow \citep{kanagawa2018radial}, who relate the type-II migration timescale with the type-I as 

\begin{equation} \label{eq:migII}
    \tau_{\mathrm{mig,II}}= \frac{\Sigma_{\mathrm{un,p}}}{\Sigma_{\mathrm{min}}}\tau_{\mathrm{mig,I}}
\end{equation}
where $\Sigma_{\mathrm{un,p}}$ is the surface density of the unperturbed disc while $\Sigma_{\mathrm{min}}$ corresponds to the surface density at the bottom of the gap. They are related as \citep{duffell2013gap,fung2014empty,kanagawa2015formation}

\begin{equation} \label{eq:sigma_ratio}
    \frac{\Sigma_{\mathrm{un,p}}}{\Sigma_{\mathrm{min}}} = 1+ 0.04K
\end{equation}
with 
\begin{equation} \label{eq:kapa}
    K=\left(\frac{M_{\mathrm{pl}}}{{M}_{\star}}\right)^{2}\left( \frac{H}{r}\right)^{-5}\alpha^{-1} \ .
\end{equation}

 To account for the damping of the orbital eccentricity and inclination due to the gravitational interaction between the disc and the planets, we follow \citep{cresswell2006evolution} and \citep{cresswell2008three} who found that the timescale of the eccentricity damping is best described by 
 \begin{equation} \label{eq:taf_ec}
 \begin{aligned}
    \tau_{\mathrm{e}} = \frac{\tau_{\mathrm{wave}}}{0.780} & \left[1  - 0.14 \left( \frac{e}{h} \right)^{2} + 0.66 \left(\frac{e}{h} \right)^{3}\right.
    \\ &  
    \left. + 0.18\left( \frac{e}{h}\right)\left(\frac{i}{h}\right)^{2}\right]
    \end{aligned}
\end{equation}
while the timescale for the inclination damping is given by
\begin{equation} \label{eq:taf_incl}
    \begin{aligned}
    \tau_{\mathrm{i}} = \frac{\tau_{\mathrm{wave}}}{0.544}  & \left[1 - 0.30 \left(\frac{i}{h}\right)^{2} + 0.24 \left(\frac{i}{h}\right)^{3}\right.
     \\ &  \left.
     + 0.14 \left(\frac{e}{h}\right)^{2}\left(\frac{i}{h}\right)\right]
    \end{aligned}
\end{equation}
 where $\tau_{\mathrm{wave}}$ is the characteristic time of the orbital evolution \citep{tanaka2004three} 
 \begin{equation} \label{eq:taf_wave}
 \tau_{\mathrm{wave}} = \left(\frac{M_{\star}}{M_{\mathrm{pl}}}\right) \left( \frac{M_{\star}}{\Sigma_{\mathrm{g,pl}}\alpha_{\mathrm{pl}}^{2}}\right) h^{4}\Omega_{\mathrm{K,pl}}^{-1} \ .
\end{equation}

These timescales for eccentricity and inclination damping are valid for low mass planets that undergo type-I migration. For planets that open gaps and start to migrate in the type-II regime, we follow the classical K-damping prescription \citep{lee2002dynamics}. The damping rates of eccentricity and inclination for massive planets that migrate in type-II regime are

\begin{equation} \label{eq:damp_rates_typeII}
\dot{e}/e = -K_{\mathrm{damp.}} |\dot{a}/a|,  \hspace{0.5cm}
\dot{i}/i = -K_{\mathrm{damp.}} |\dot{a}/a|
\end{equation}

Here we use $K_{\mathrm{damp.}}$=5. Higher $K_{\mathrm{damp.}}$ values result in higher damping rates. For a better understanding of how different $K_{\mathrm{damp.}}$ values affect the formation of giant planets, we refer to \cite{bitsch2020eccentricity} where they investigated the influence of different damping rates on the eccentricity and inclination of growing and migrating giant planets. 

\begin{figure}
  \centering
   \includegraphics[width=\hsize]{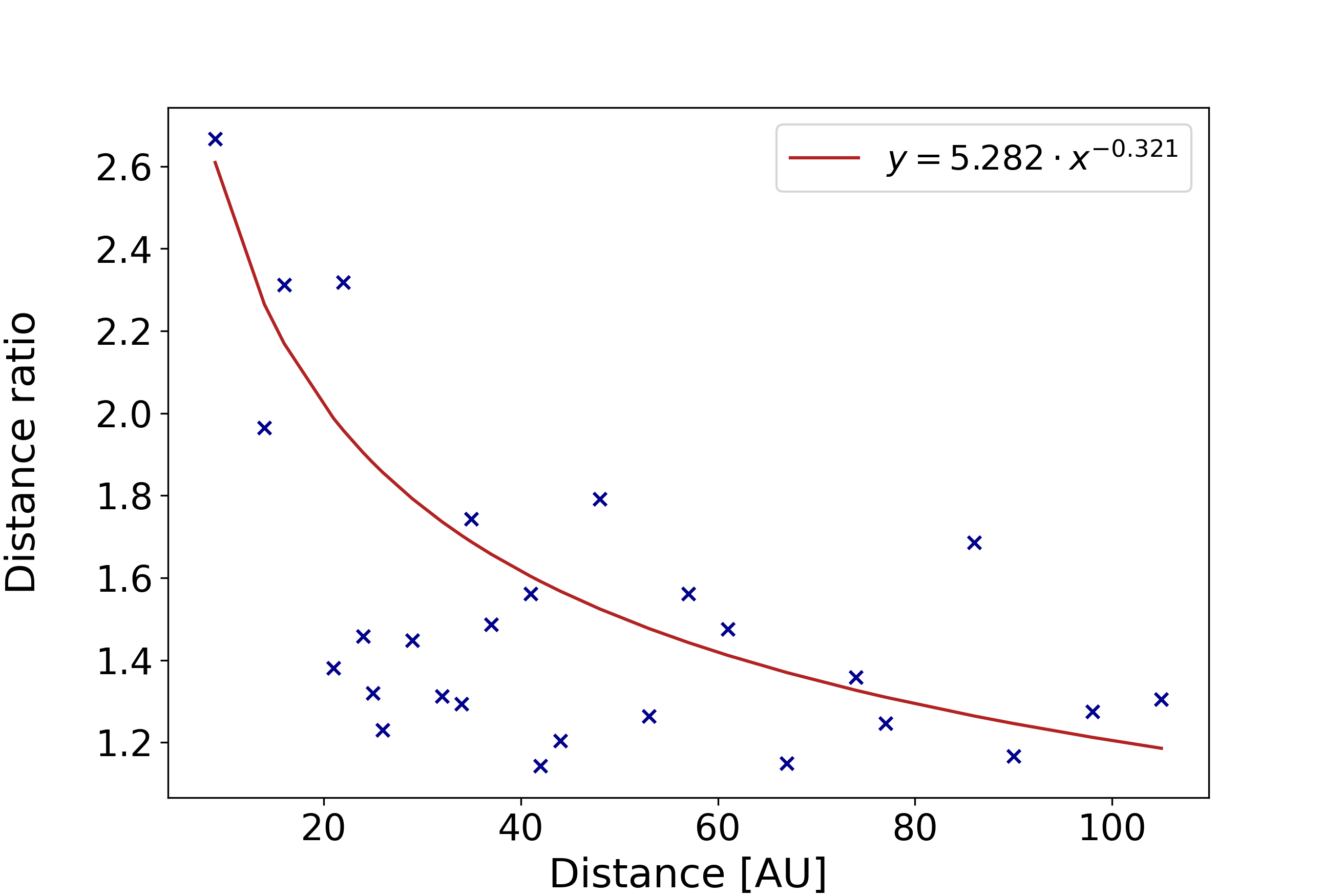}
   \caption{Power law fit between the position of the gaps observed by ALMA 1.25mm continuum images (DSHARP) \citep{huang2018disk} and the distance between them. }
    \label{fig:power_law}
\end{figure}
\begin{figure*}[h!]
\centering
   \includegraphics[width=17cm]{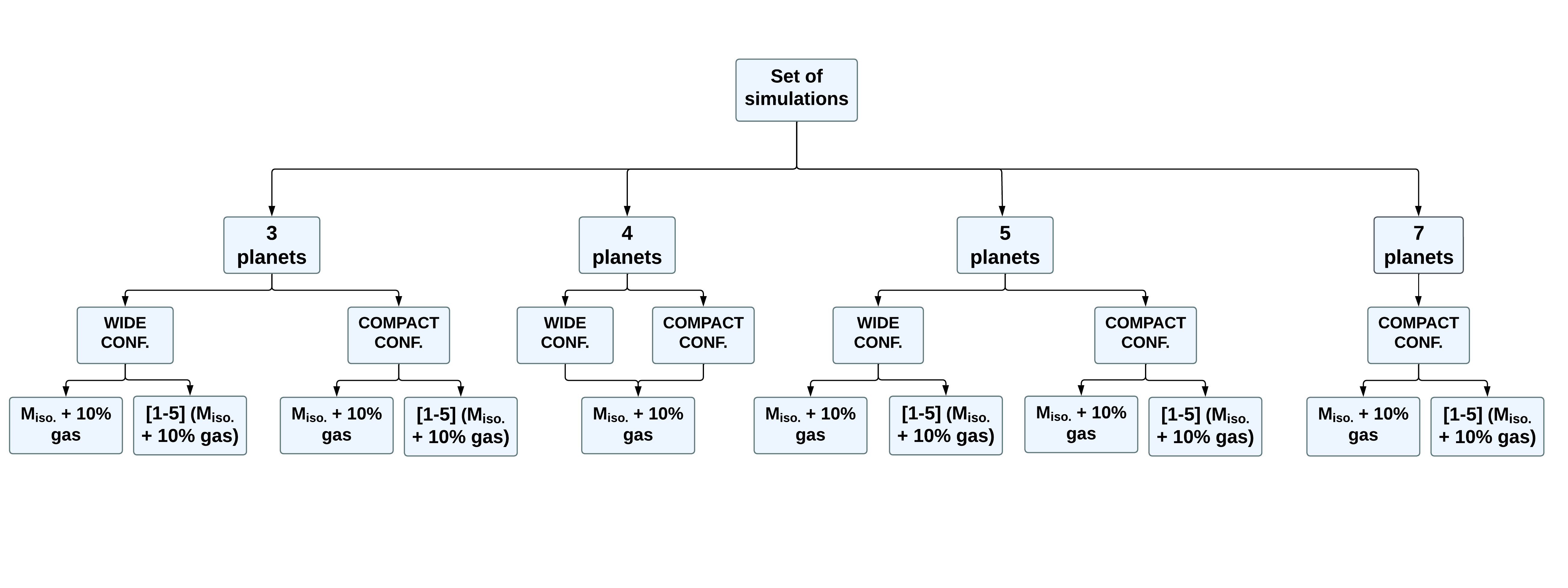}
    \caption{All the different setups we used for our simulations. These setups were tested for  three $\mathrm{\alpha}$ values (see Table \ref{table:disc}).}
    \label{fig:simulations}
\end{figure*}

\subsection{Initial Conditions} \label{initial_conditions}
We model the evolution in the case where we have three, four, five and seven putative planets. For the position of these planets in each case, we took the annular substructures of the 18 disc systems from the Disc Substructures at High Angular Resolution Project (DSHARP) \citep{huang2018disk} and fit a power law between the position of these substructures and the distance between them. The result of the fit is shown in Fig. \ref{fig:power_law}. \textbf{Considering an inner planet located at 10 AU, this yields initial reference locations at 10, 25, 47, 72, 97, 118, 135, 147 and 156 AU. }
To account for the uncertainties in the initial position of the planets, we ran simulations for two different configurations, one where the planets are located in a wide configuration and another where they are in a compact configuration. \textbf{The wide configuration for the 3 planet case harbors initial positions at 10, 47 and 97 AU while the compact configuration has initial planetary positions at 25, 47 and 72 AU. For the four planet case the wide configuration gives starting positions at 10, 47, 97 and 135 AU and for the compact configuration at 25, 47, 72 and 97 AU.} To further address any uncertainties, we also add an extra 5$\%$ of deviation for the individual starting positions of our planets. 
 For this planet distribution, we assume that the planets have just reached the pebble isolation mass (Eq. \ref{eq:Miso}) and they start to accrete gas. We chose to test two distinct simulation set up for their initial mass. The first case is where we considered that the planets have already accreted 10$\%$ of the gas on top of the pebble isolation mass. As second case we have a mass distribution chosen randomly between 1 and 5 times of the pebble isolation mass plus 10$\%$ of gas. Finally, the disc lifetime $\mathrm{t_{disc}}$ were randomly selected from the range of [3-10]Myr (Table \ref{table:disc}) although we evolve the systems until 100Myr to study also their evolution after the dissipation of the disc. \\
\par
As in \citep{bitsch2019formation} the initial inclinations and eccentricities of the planets were distributed uniformly with inclinations between $0.01^{\circ} - 0.05^{\circ}$ and eccentricities between 0.001 - 0.01. The other orbital angles were randomly selected between $0^{\circ} - 360^{\circ}$. We used three different $\alpha$-viscosities, $\alpha = 10^{-3}, 3.16 \cdot 10^{-4}$ and $10^{-4}$. Higher values of viscosity result in higher migration and accretion rates. Higher $\alpha$ values result in a shallower gap because of eq \eqref{eq:kapa}, and the higher accretion rate originates from eq \eqref{eq:M_dotdisc}.
For each $\alpha$ value, we ran 50 simulations with different initial conditions for the positions of the planets. Fig. \ref{fig:simulations} shows a tree diagram illustrating all the different sets of our simulations.

\section{Results} \label{results} 
In this section we present the outcomes of our simulations regarding the distribution of mass, semi-major axis, and eccentricity for different planetary systems.  We organize our results by considering the number of planets in each system. We first show results for the three-planet system in Sect. \ref{res:3pl} and next we present our results for the four-planet system in Sect. \ref{res:4pl}. We then investigate systems consisting of five and seven planets in Sect. \ref{res:5-7pl}.

\subsection{Three Planet System} \label{res:3pl}
\begin{figure*}[h!]
	\centering
	\begin{minipage}{\columnwidth}
		\centering
		\includegraphics[width=\hsize]{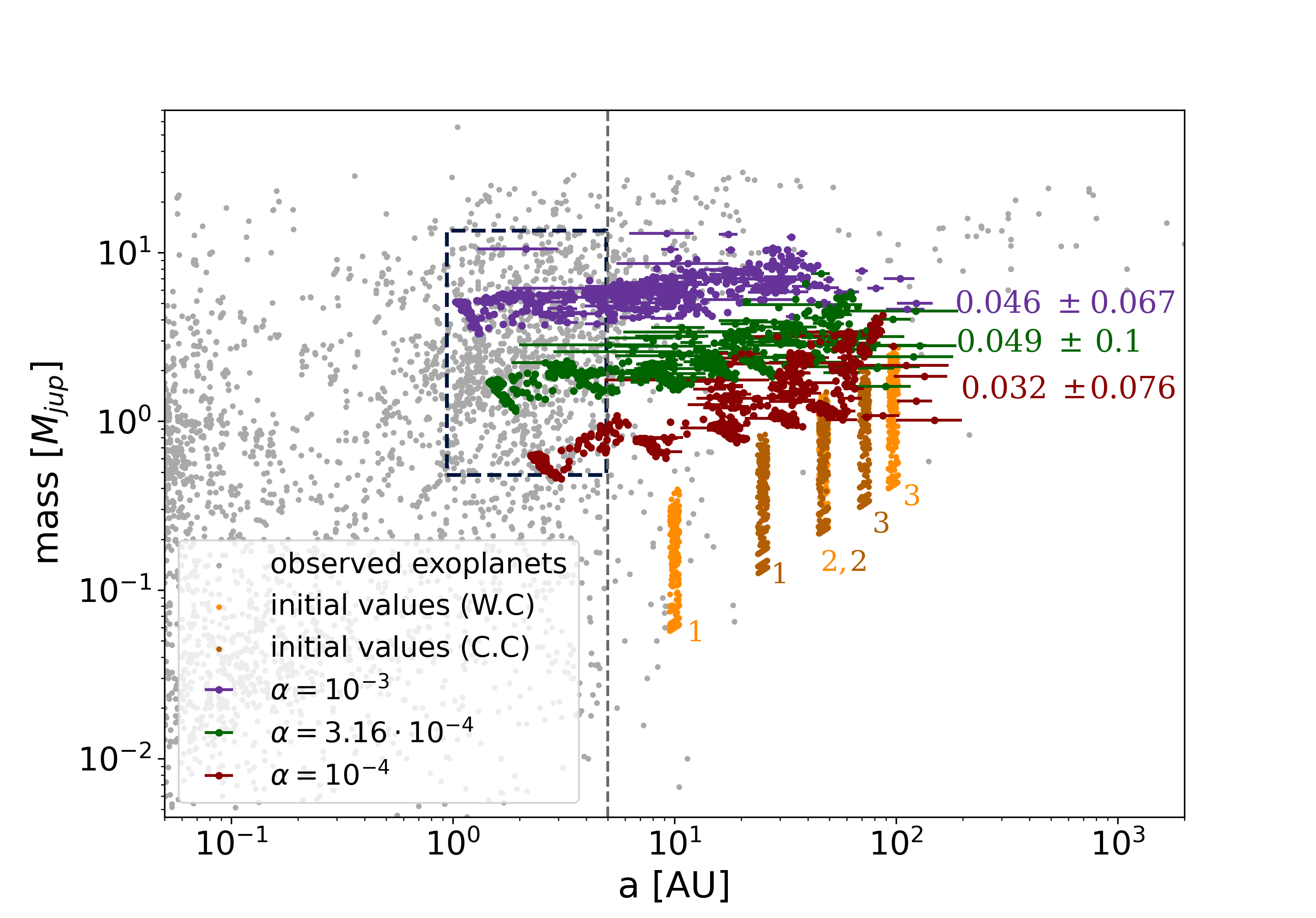}
	\end{minipage}%
	\begin{minipage}{\columnwidth}
		\centering
		\includegraphics[width=\hsize]{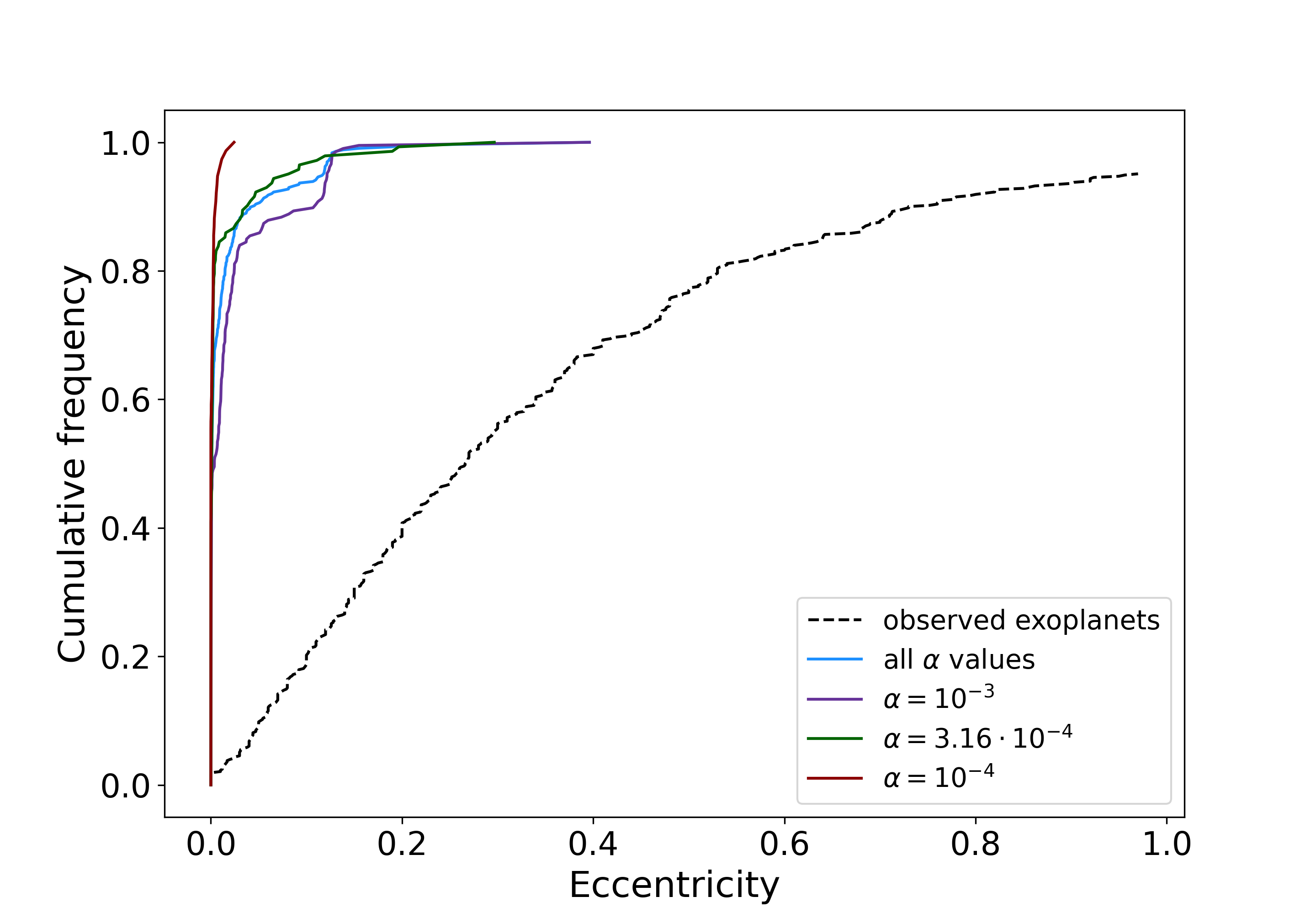}
       
	\end{minipage}
 \caption{The left figure shows the comparison between the observed giant exoplanet population (gray dots) and all the simulations for the three- planet systems. The initial values of these planets are denoted as orange for the case where the planets  located in a wide configuration while brown dots denote planets located in a compact configuration, where the small numbers indicate the position of the first, second, and third planet. Purple, green and red data points symbolize the simulations for the different $\mathrm{\alpha}$, $10^{-3}$, $3.16 \cdot 10^{-4}$ and $10^{-4}$ respectively, whereas the coloured numbers shows the mean value and the standard deviation of the eccentricity for each case for all the simulated three- planet systems. The horizontal lines refers to the perihelion and aphelion of the planet $\mathrm{a(1 \pm e)}$. The black dashed vertical line represents the current Radial Velocity (RV) detection limit at 5 AU. Right figure shows the cumulative frequency of the eccentricity for planets up to 5 AU and mass between $ \mathrm {0.5 \leq mass [\mathrm{M_{Jup}}] \leq 13 }$. We also note that these plots include both cases for the initial mass (fixed and random value). }
 \label{fig:3pl}
\end{figure*}

Fig. \ref{fig:3pl} shows the final locations and masses of the simulated three-planet systems on a mass vs.\ semi-major axis plot after 100Myr of integration time (left) as well as their eccentricity distribution (right). This plot includes both configurations for the position of the planets and also both configurations for their initial mass. Orange dots refer to the initial conditions where the planets are located in a wide configuration whereas the brown dots refer to the more compact configuration. The different colours represent simulations of different $\alpha$ viscosity values. The horizontal line segments give an indication of how eccentric our simulated planets are as they extend from the perihelion $a(1 - e)$ to aphelion $a(1 + e)$. To provide a clearer indication of the final eccentricity, we have included the average eccentricity values for the three sets of different $\alpha$ values. These values are represented by colored numbers. 

\par In general the planets experience gas accretion and migration towards the central star, as evident by the increased mass and inward migration. As we mentioned earlier, we expect to have higher migration and accretion rates for the higher values of the $\alpha$-viscosity  $10^{-3}$. This is exactly what we see in the left panel of Fig. \ref{fig:3pl}. In Table \ref{table:results} we show the outcomes of our simulations for the different $\alpha$ viscosity values. For higher values of the $\alpha$ viscosity the accretion of gas and the orbital migration is more efficient. On the other hand, simulations with the lowest value $\mathrm{\alpha} = 10^{-4}$ exhibit relatively slower rates of migration and accretion, while for the intermediate $\alpha$ value, we also find intermediate planetary masses. These results are  consistent with those of \citep{muller2022emerging}, where they investigated a specific target, the disc of HD 163296, by modelling the evolution of the three planet candidates.

\par To compare our results with observations, in Fig.\ \ref{fig:3pl} we also plotted the current observed exoplanets (gray dots). The data we used were obtained from the NASA Exoplanet Archive and were taken on 3 of February in 2023 (\url{https://exoplanetarchive.ipac.caltech.edu/}). The black dashed vertical line represents the current Radial Velocity (RV) detection limit at 5AU. \textbf{This limit corresponds to the time limit of how long observations would need to be done, so it is a mass - independent limit which is valid for the mass range considered here \citep{wright2012exoplanet}.} Only a small fraction of the synthesised planets migrate into the inner disc and none interior to 1 AU. This already hints to the fact that planets such as those assumed to be originating from gaps in protoplanetary discs might not be the sole source of the currently observed giant planet population. In addition, there is a considerable number of planets that lie beyond the RV detection limit of 5 AU. When these detection limitations are removed, we will be able to confirm or exclude the presence of such wide-orbit giants.

\par In the right panel of Fig.\ \ref{fig:3pl}, we present the cumulative distribution of the eccentricities, separately for each $\mathrm{\alpha}$-viscosity value, as well as for the combined dataset (blue line). We also show the cumulative distribution of eccentricity for the observed exoplanet sample as a dashed black line. The dataset is limited to include only those simulated planets with semi-major axis of  $1 \leq a~ \mathrm{[AU]} \leq 5$ and masses within the range of $0.5 \leq m~ \mathrm{[M_{Jup}]} \leq 13 $. These values are marked by the black box in the left panel of Fig. \ref{fig:3pl}. We also limit the observed exoplanet sample to include only systems for which the mass of the central star is between $0.7 \leq m~ \mathrm{[M_{\odot}]} \leq 1.3$ in order to have a fair comparison with our simulated systems where we have assumed the central star mass of $\mathrm{m = 1 M_{\odot}}$. It is clear that the eccentricity distribution of our simulated planets is too low  (last column of Table \ref{table:results}) compared to the eccentricity distribution of the observed giant planets (mean value of 0.302).

\subsection{Four Planet System} \label{res:4pl}

\begin{figure*}
	\centering
	\begin{minipage}{\columnwidth}
		\centering
		\includegraphics[width=\hsize]{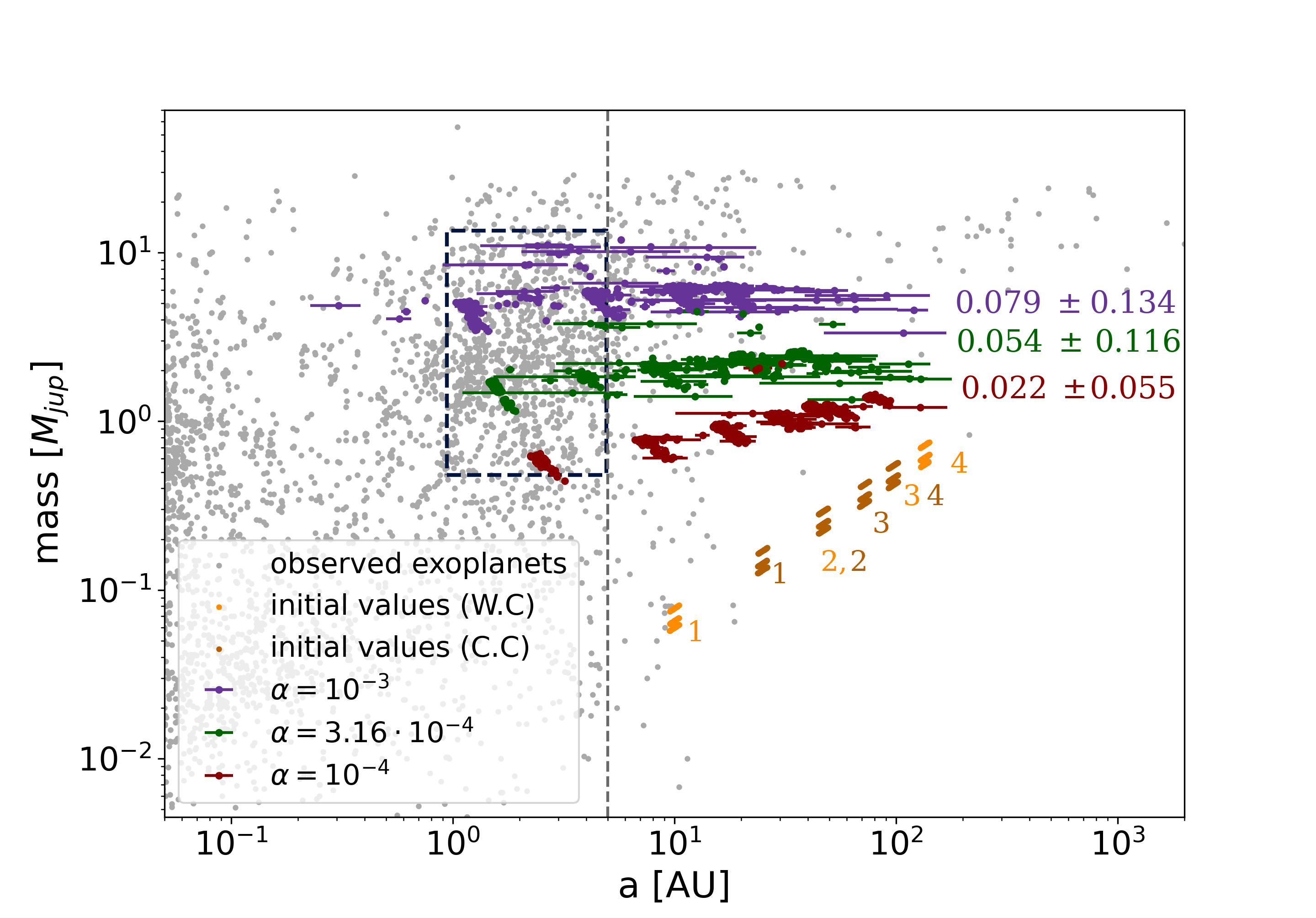}
	\end{minipage}%
	\begin{minipage}{\columnwidth}
		\centering
		\includegraphics[width=\hsize]{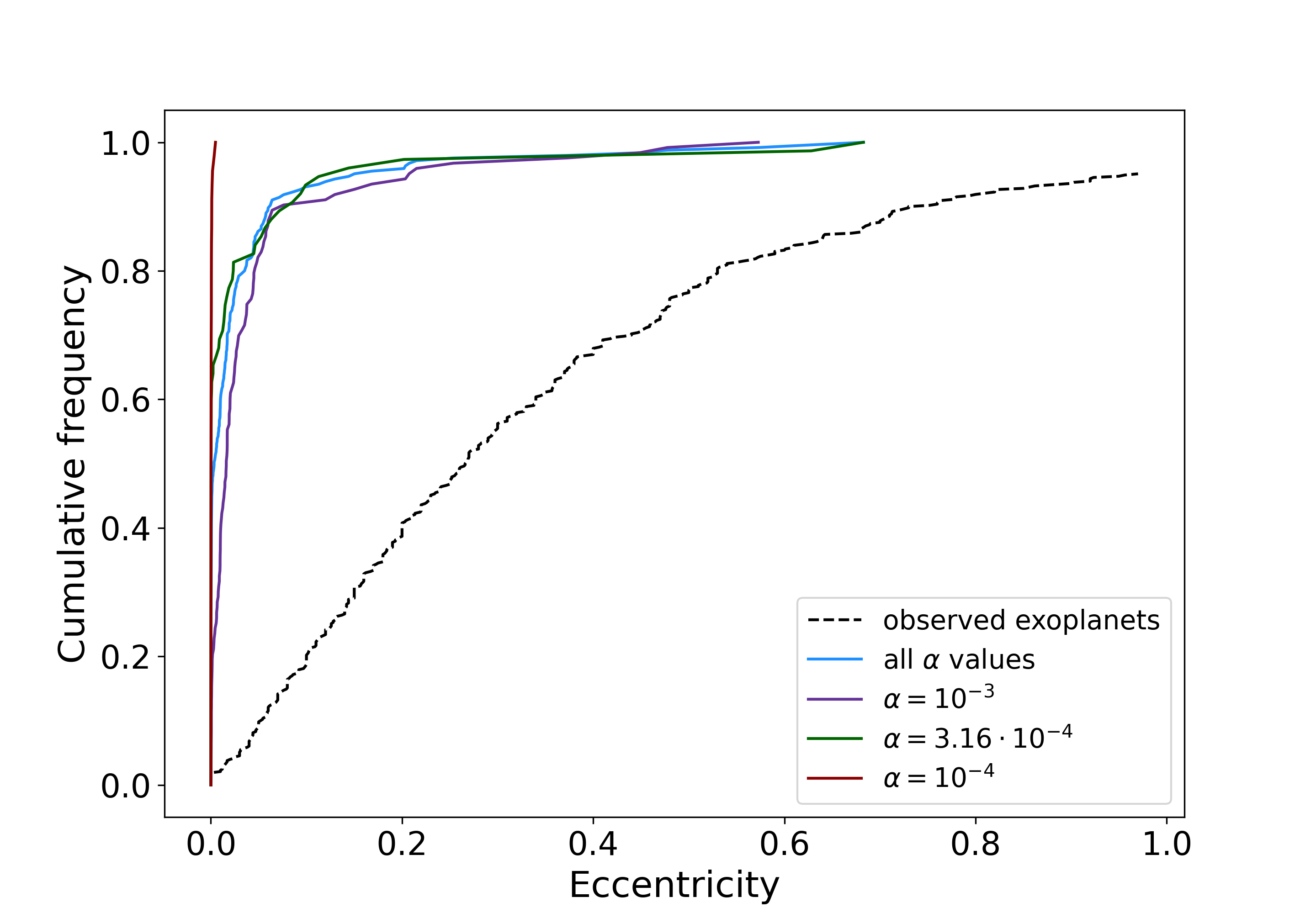}
       
	\end{minipage}
 \caption{Similar to Fig. \ref{fig:3pl}, but for  systems with initially four planets. }
\label{fig:4pl}
\end{figure*}

\begin{table}[h!]
\caption{Mean value and standard deviation of the number of surviving planets for the three and four-planet system after 100Myr of integration time.}            
\label{table:surv_pl}      
\centering                         
\begin{tabular}{c c c c}        
\hline\hline                
system & mean value & $\sigma$ \\   
\hline                        
   3 planets & 2.83 & 0.47  \\      
   4 planets & 3.50 & 0.98     \\ 
\hline                                 
\end{tabular}
\end{table}

Fig \ref{fig:4pl} shows the mass vs.\ semi major axis plot of the simulated four-planet systems (left) and their eccentricity distribution (right) after 100 Myr of evolution. The plots are analogous to Fig. \ref{fig:3pl}. We also note that for the initial mass in these systems we only tested the case where $m = M_\mathrm{iso} + 10\%$ gas. If we increase the mass any further, the systems would become unstable on a very short time scale, see Sect.\ \ref{dis:unstable} for further discussion on this issue.

\par In the left panel of Fig.\ \ref{fig:4pl}, we see that the planets experience a similar evolutionary pattern to the three-planet system. The planets undergo gas accretion and migration towards the central star, and the efficiency of these processes increases as the $\mathrm{\alpha}$ viscosity increases. We present the final semi-major axis, mass and eccentricity in Table \ref{table:results} for the different $\alpha$ viscosity values as well as for the combined dataset.
\par The right panel of Fig.\ \ref{fig:4pl}, shows the cumulative distribution of eccentricities for our simulated four-planet systems after 100Myr of integration time. We compared to the same set of planets regarding their mass, distance and stellar host mass as for the three-planet case. For the lowest $\alpha$  viscosity value, $\alpha = 10^{-4}$, the eccentricities are too low, whereas for higher viscosities for which the rates of gas accretion and migration are more effective, the final eccentricities are a little bit larger. However, our simulated eccentricities are clearly too low to match the observations. The precise values of the eccentricities for the limited dataset are presented in the last column of Table \ref{table:results}.
\par Compared to the three-planet system, the distribution of mass and semi-major axis of our simulated planets are very similar to each other. On the other hand, the eccentricity distribution for systems consist of four planets is slightly higher - two times higher - than the systems with 3 planets, but still too low to match the observations. During the gas disc phase, the eccentricity of the planets are damped because of their interaction with the disc, resulting in low eccentricities. After the dissipation of the disc, the planets can interact with each other leading to an increase in their final eccentricity. However, the distances between the objects are too large to allow efficient scattering events. Table \ref{table:surv_pl} shows the average number of surviving planets at the end of our simulations. We see that for systems with initially four planets the occurrence of scattering events between planets is slightly higher. Hence, it is reasonable to obtain higher mean eccentricity for the four-planet systems.

\begin{figure*}[h!]
	\centering
	\begin{minipage}{\columnwidth}
		\centering
		\includegraphics[width=\hsize]{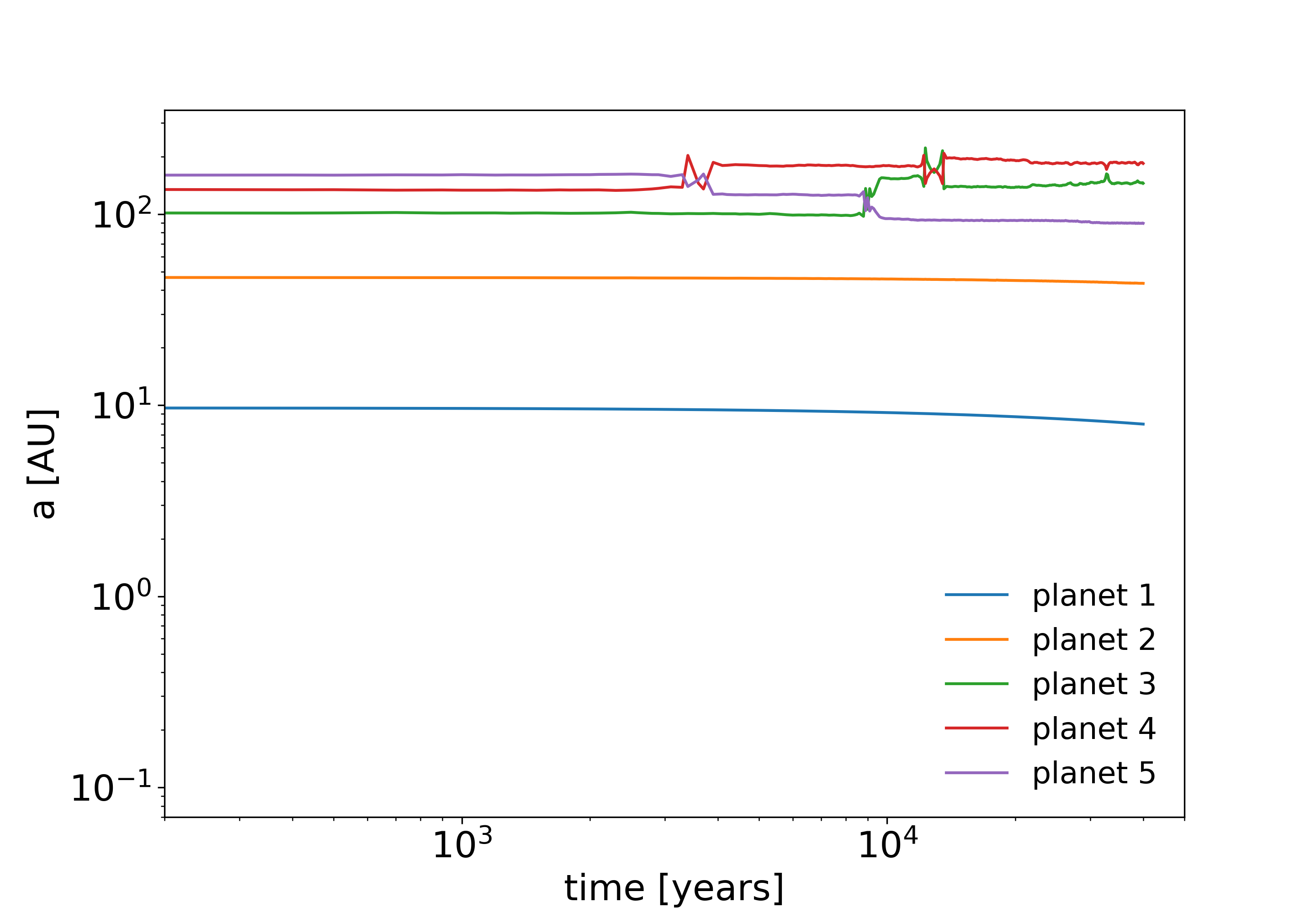}
	\end{minipage}%
	\begin{minipage}{\columnwidth}
		\centering
		\includegraphics[width=\hsize]{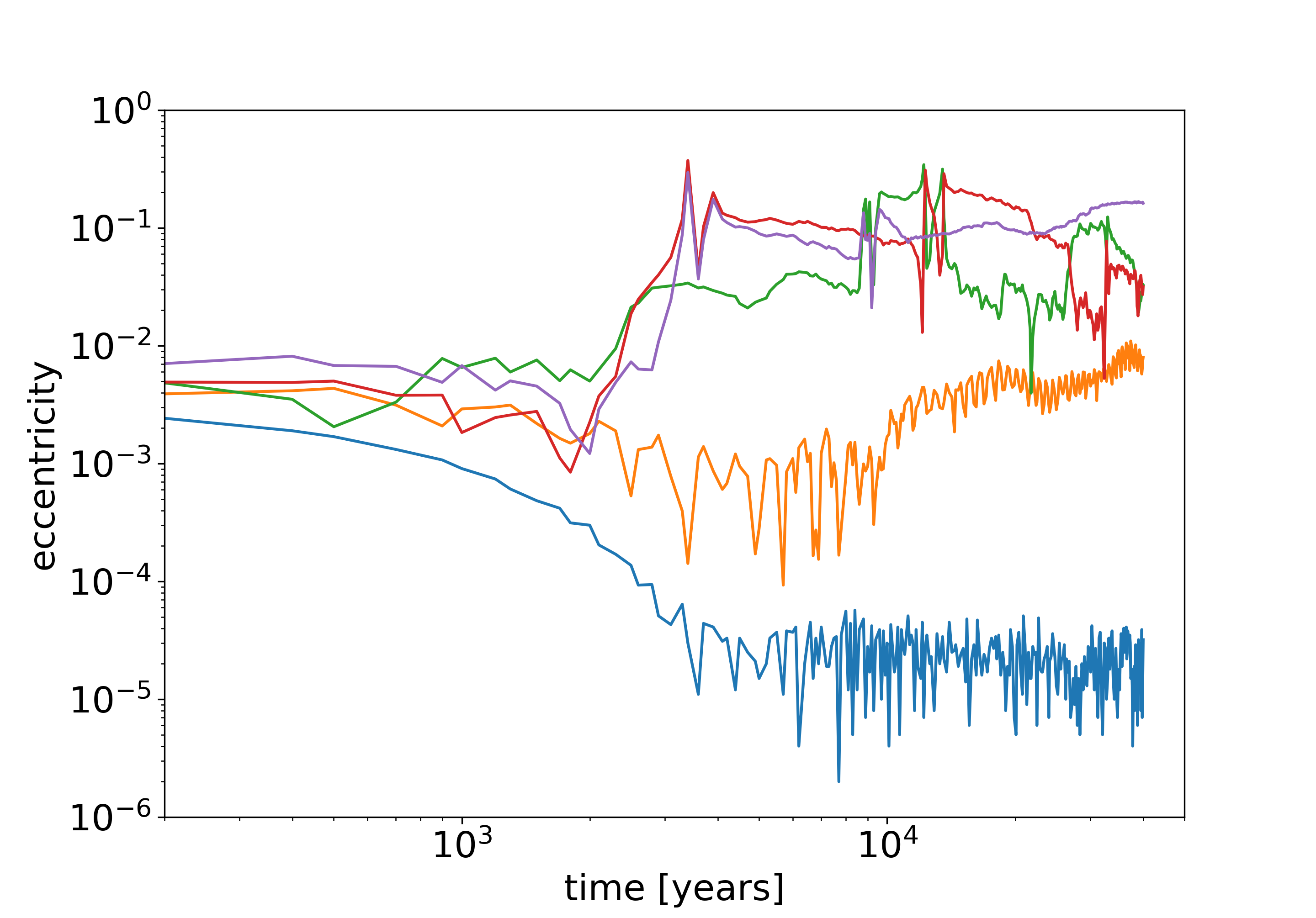}
       \end{minipage}%
\end{figure*}
\begin{figure*}[h!]
	\centering
	\begin{minipage}{\columnwidth}
		\centering
		\includegraphics[width=\hsize]{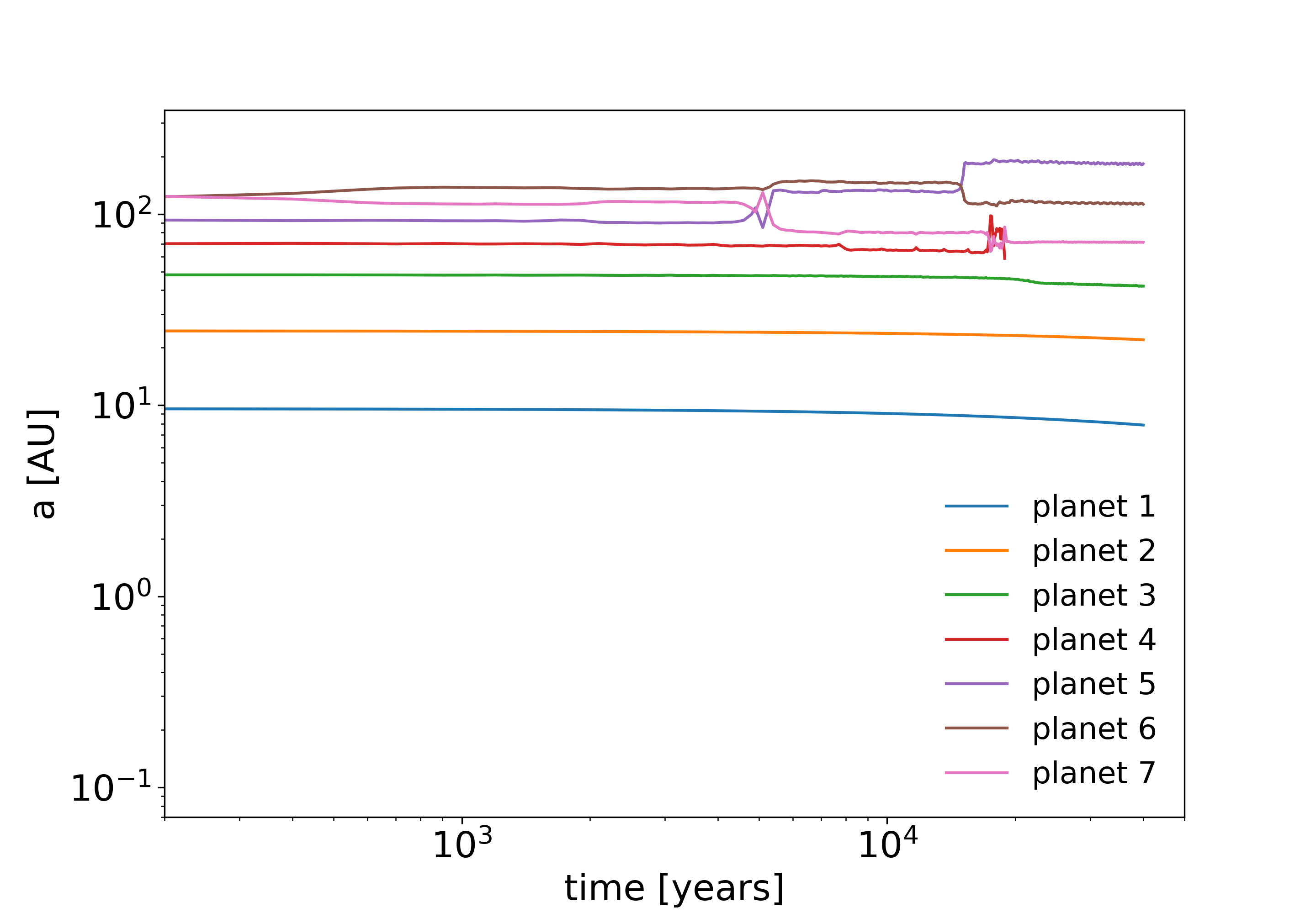}
	\end{minipage}%
	\begin{minipage}{\columnwidth}
		\centering
		\includegraphics[width=\hsize]{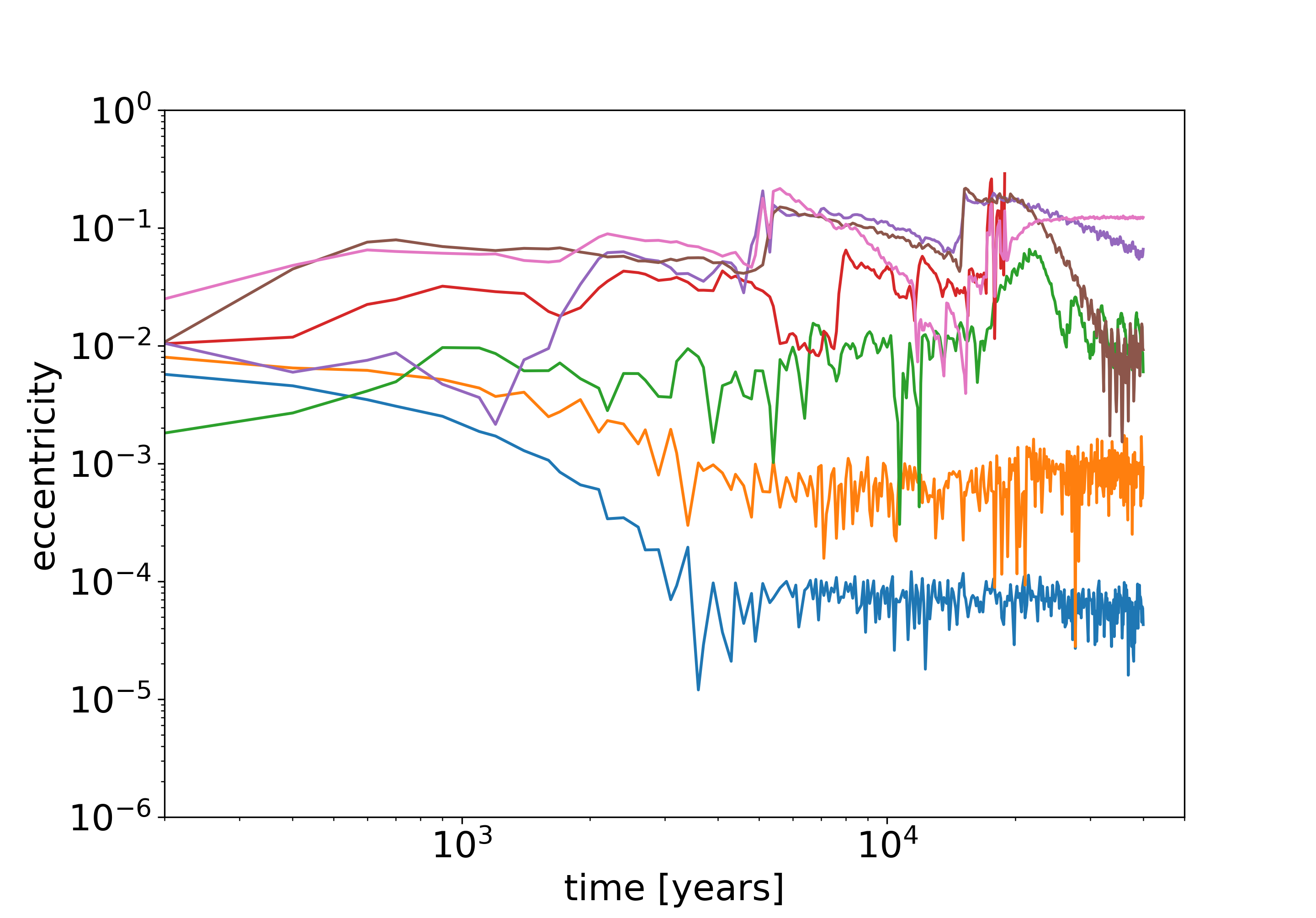}
       \end{minipage}%
 \caption{Evolution of two different individual systems for the 5 (top) and 7 (bottom) planet system. The plots shows the evolution of the semi-major axis (left) and the eccentricity (right) over time.}
 \label{fig:individual_systems}
\end{figure*}

\begin{table*}[h!]
\caption{Outcomes of the simulations of the three and four planet systems. The results correspond to the total dataset. We also show the mean eccentricity for the limited dataset (right figure of Fig. \ref{fig:3pl} and Fig. \ref{fig:4pl}) in the last column.} 
\label{table:results} 
\centering 
\begin{tabular}{c c c c c} 
\hline\hline 
system & $\alpha$-viscosity  & mass $(M_{jup})$ & eccentricity & eccentricity (lim. dataset) \\
\hline 
\\ 
3 planets & $10^{-3}$ & 5.921 $\pm$ 1.392 & 0.046 $\pm$ 0.067 & 0.023 $\pm$ 0.046\\
 & $3.16 \cdot 10^{-4}$ & 2.417 $\pm$ 0.885 & 0.049 $\pm$ 0.100 & 0.013 $\pm$ 0.039\\
 & $10^{-4}$ & 1.320 $\pm$ 0.743 & 0.032 $\pm$ 0.076 & 0.002 $\pm$ 0.004\\
 & $10^{-4} - 10^{-3}$ &  3.175 $\pm$ 2.210 & 0.042 $\pm$ 0.083 & 0.016 $\pm$ 0.040\\
\\
4 planets & $10^{-3}$ & 5.584 $\pm$ 1.281 & 0.079 $\pm$ 0.134 & 0.044 $\pm$ 0.091\\
 & $3.16 \cdot 10^{-4}$ &  2.099 $\pm$ 0.439 & 0.054 $\pm$ 0.116 & 0.034 $\pm$ 0.109\\
 & $10^{-4}$ &  0.993 $\pm$ 0.272 & 0.022 $\pm$ 0.055 & 0.0005 $\pm$ 0.0009\\
 & $10^{-4} - 10^{-3}$ &  2.698 $\pm$ 2.054 & 0.050 $\pm$ 0.107 & 0.033 $\pm$ 0.090\\
\\
\hline 
\end{tabular}
\end{table*}

\subsection{Five \& Seven Planet system} \label{res:5-7pl}

\begin{table}
\caption{Distance between two planets in mutual Hill radius at the beginning of the integration time $t=0$. For the mass of the planets we assumed $m=M_{\mathrm{iso}}+10\%$.} 
\label{table:distance_hill} 
\centering 
\begin{tabular}{c c c c} 
\hline\hline 
system & configuration & planet neighbors &  distance  \\
\hline 
\\ 
5 planets & wide & 1-2 & 26.052 \\
& & 2-3 & 10.618 \\
& & 3-4 & 4.362 \\
 &  &  4-5 & 1.878  \\
\\
5 planets & compact & 1-2 & 11.383 \\
& & 2-3 & 6.773 \\
& & 3-4 & 4.312 \\
& & 4-5 & 2.657  \\
\\
7 planets & compact & 1-2 & 19.659 \\
& & 2-3 & 11.383 \\
& & 3-4 & 6.773 \\
& & 4-5 & 4.312 \\
& & 5-6 & 2.657 \\
& & 6-7 & 1.745 \\

\\
\hline 
\end{tabular}
\end{table}

For the five planet system we studied two configurations for the initial position of the planets whereas for the seven planet system we studied only systems in compact configuration. For both systems we also tested two configurations for the initial mass. Fig.\ \ref{fig:individual_systems} shows two individual systems of five (top) and seven planets (bottom). Each planet is represented by different colour. The right panel shows the evolution of the eccentricity over time and the left panel shows the evolution of the semi-major axis. We integrate each system for 40Kyr. From Fig.\ \ref{fig:individual_systems} we see that both systems are initially unstable leading to scattering between the outermost planets. 
We also note that the individual systems presented in Fig.\ \ref{fig:individual_systems} are not the exception but rather a common outcome of these simulations. All the systems of five and seven planets and different simulation set-up we studied exhibit this instability early on. 
\par We can explain these initial instabilities to some level, by examining the stability of planetary systems via the Hill criterion. \citep{gladman1993dynamics} studied the dynamics of two small planets orbiting the Sun. He found that in order to have stable systems, the difference of the initial semi-major axis between two planets measured in mutual Hill radius ${R_{\mathrm{H}} =  [(m_{1} + m_{2})/ 3M_{\odot}]^{1/3} [(a_{1} + a_{2})/2] }$, must be greater than $\mathrm{2 \sqrt{3}}$. In Table \ref{table:distance_hill} we show the distance between two planets in mutual Hill radius at $t=0$yr. These distances were calculated by assuming that the mass of the planets is $m = M_\mathrm{iso} + 10\%$. We see that the distance between the outermost planets in both systems is much lower than the critical value of $2 \sqrt{3}$, explaining why the systems become unstable. This instability is clearly not prevented by the damping forces of the disc.

\section{Discussion} \label{discussion}

\begin{table}[h!]
\caption{Mean distance in mutual Hill radius for the three and four planet systems for different $\alpha$-viscosity values after 100Myrs of integration time. The results correspond to both configurations for the positions of the planets (wide \& compact) and both configurations for their initial mass (fixed \& random).} 
\label{table:hill_3_4}
\centering                         
\begin{tabular}{c c c c}        
\hline\hline                
system & $\alpha$ & planet neighbors & distance \\   
\hline         
\\
 3 planets  & $10^{-3}$ & 1-2 &  7.07\\
   & & 2-3 & 4.79\\
  \\
  & $3.16 \cdot 10^{-4}$ & 1-2 & 9.79\\
  & & 2-3 & 6.29 \\
  \\
  & $10^{-4}$ & 1-2 & 12.63 \\
  & & 2-3 & 7.35 \\ 
  \\
4 planets  & $10^{-3}$ & 1-2 & 6.90 \\
  & & 2-3 & 4.75 \\
  & & 3-4 & 3.80 \\
  \\
   & $3.16 \cdot 10^{-4}$ & 1-2 & 9.97 \\
  & & 2-3 & 6.37 \\
  & & 3-4 & 4.80 \\ 
  \\
   & $10^{-4}$ & 1-2 & 13.95 \\
  & & 2-3 & 8.18 \\
  & & 3-4 & 6.03 \\
  \\
\hline                                 
\end{tabular}
\end{table}

The presence of rings and gaps in the dust emissions of protoplanetary discs is often attributed to young planets and more decisive evidence of the presence of planets has been additionally mustered via gas kinematics \citep{zhang2015evidence,flock2015gaps,gonzalez2017self,takahashi2016origin,pinilla2012ring,long2018gaps}. Still, it is not clear at the present day whether these substructures should be attributed to planets as a general rule, nor whether each dip in the dust emission is to be associated with one single planet \citep{dong2017multiple,bae2017formation}. Another complementary but related question concerns the fate of such growing planets in the context of planet formation theory and the currently observed exoplanet sample. In particular, we can ask whether these putative young forming planets represent the progenitors of the currently observed giants in the sample: if yes, we can draw information on the formation history of the currently observed planets and better inform planet formation theory; if not, we should either expect a yet-undetected population of planets, or, if future observations rule out such an unseen population, put into question our formation scenarios of planets forming efficiently in gaps. To this end, in this section we systematically compare the outcome of our simulations of growing and migrating multi-planetary systems with the observed giant exoplanets.

\subsection{Stable systems} \label{dis:stable}

We focus here on the outcomes concerning the systems with initially three and four planets. Compared to the observed eccentricity distribution our simulations clearly show too low eccentricities of the surviving planets. A reason for this outcome could be that the planets do not interact much with each other, avoiding scattering events resulting in low eccentricities. The average number of the surviving planets after 100Myr of integration time is 2.83 for the three planet system and 3.50 for the four planet system. 

\textbf{The further evolution beyond 100 Myr would most likely not affect the eccentricity and semi-major axis distribution of our systems, because the distances in units of mutual Hill radii after 100 Myr of evolution are very large indicating that scattering events should be rare.}

\citep{raymond2009planet} studied the eccentricity evolution of a three-planet system located initially in an unstable configuration with a separation of $4-5R_{\mathrm{H}}$. They found that scattering among planets in this unstable configuration result to an eccentricity distribution that agrees with observations. Table \ref{table:hill_3_4} shows the distances of our simulated planets in mutual Hill radius for the different $\alpha$ viscosities after 100Myr. These values correspond to both configurations for the initial position, compact and wide configuration, of the planets and for both configurations regarding the initial mass for the three planet system and for $m=M_{\mathrm{iso}} +10\%$ for the four planet system. In our simulations, the surviving planets have a larger separation than in \citep{raymond2009planet}, indicating why our simulations do not produce efficient scattering events and thus reflect a population of low eccentricity planets. 

A way of increasing the excitation of planetary systems is the presence of leftover planetesimals. At later stages of the disc lifetime where the gas surface density drops, the dust-to-gas ratio goes up, which is a favourable condition for planetesimal formation \citep{carrera2015form}; moreover planetesimal-disc-driven instabilities are believed to have sculpted the orbital distribution of the giant planets in the outer solar system, increasing the eccentricities by a factor of 10  \citep{nesvorny2018dynamical,tsiganis2005origin}. Another possibility is the presence of planets in the inner disc. \citep{bitsch2020eccentricity,bitsch2023giants} studied the eccentricity distribution of protoplanetary embryos located in the inner disc by running N-body simulations. They found that the simulations that match best the observed eccentricity distribution of giants are those where the planets interact with each other. However, this does not necessarily rule out the possibility that planets can also exist in the outer disc. 
Injecting the planets according to the gaps' positions in the currently available observations, we limited our study only to planets located in the outer disc, due to observational biases. 
Indeed, the inner part of the disc is not resolved with the current sensitivities and so we do not observe many gaps and rings in that region. One exception is the gap located at 1AU in the TW Hya disc \citep{andrews2016ringed}.

\subsection{Unstable systems} \label{dis:unstable}

For all the different initial conditions we tested, we found that systems with five and seven planets become unstable immediately, resulting in scattering events and increased eccentricities of the surviving planets, see Fig.\ \ref{fig:individual_systems}.

\citep{lega2013early} studied the planet-planet and the planet-disc interaction in a three planet system with hydrodynamical simulations. They found that when the planets reach Jupiter mass the systems become unstable. This instability disturbs the gas surface density resulting in complex patterns, where gaps and rings are not visible. However, the dust emissions of protoplanetary discs observed in the DSHARP project do not show complex gas density distribution but rather more well defined gaps and rings. This indicates that not all gaps/rings in observed protoplanetary discs are caused by planets. 

In particular, we stress that the initial configuration for our seven planet system is very similar to the position of the seven gaps in the AS 209 system. This strongly indicates that for such discs it is not possible for all the observed gaps and rings to be caused by planets above the pebble isolation mass located at the positions of the gaps. On the other hand, planets below the pebble isolation mass cannot be responsible for the observed gap structures. Therefore, additional processes that can cause gaps/rings without necessarily assuming one forming planet for each gap are needed to understand disc substructure observations. For example, \citep{bae2017formation} showed that a single 30 Earth-mass planet in a low ($\alpha = 5 \cdot 10^{-5}$) viscosity disc can reproduce multi-ring features. \textbf{Additionally, \citep{dong2018multiple} also found that the multiple gaps in the HL Tau, TW Hya and HD 163296 can be produced by a single  planet in a low viscosity disc ($\alpha \leq 10^{-4}$). Both studies align with our findings that not each gap is explained by the presence of a planet}. Moreover, there exist various purely hydrodynamical, magneto-hydrodynamical and dust-evolution mechanisms that alone can produce substructures such as rings and gaps in protoplanetary discs \citep{rice2004accelerated,kuznetsova2022anisotropic,marcus2015zombie,flock2017radiation,bae2019ideal,li2021thresholds,flock2015gaps,suriano2018formation}

\subsection{Disc model} \label{dis:disc_model}

In this study we tested the effects of three different values of $\alpha$ viscosity within our simulations (e.g the viscosity influences the disc structure as well as the planetary migration and accretion rates), but we limited our study to a constant value of $\alpha$ with respect to time and orbital distance. If the $\alpha$ viscosity was radially dependent, with $\alpha$ increasing with distance from the star, then planets located further out in the disc would have higher migration rates. Such a radial dependence in $\alpha$ would lead to planets being located in a closer configuration and could possibly result to more planet-planet interactions. However, it seems that the viscosity in the majority of the disc might actually be quite low \citep{lesur2022hydro,delage2022steady}. 
\par 
We also note that the outcomes of our simulations are not affected by the surface density profile. In fact, we tested all the different simulation setups for another surface density profile where $\Sigma_{0}=283gcm^{-2}$. Even with the lower value of $\Sigma_{0}$ the final distribution of the planets in the mass-semi major axis plot as well as the eccentricity distribution were very similar to the distribution we presented in Sect. \ref{results}. Furthermore, the systems with five and seven planets became unstable immediately in this case as well, as we would expect.

\subsection{Implications for observations}
Another outcome of our simulations, regardless of the level of the $\alpha$-viscosity value, is the presence of far-out giant planets farther out than 5AU which is the current RV limit. Future observations will be able to either confirm or rule out the existence of such planets. In the latter scenario, it would suggest that either our formation models are incomplete, or that the putative planets located at the gaps observed in young discs are not the progenitors of the current exoplanet sample.

\section{Summary} \label{summary}

In recent years, observations of disc substructures have drawn considerable attention due to their connection to disc evolution, planet formation and their mutual influence: how the disc environment influences the growth, migration and orbital properties of the forming planets and how these forming planets affect the disc structure. In particular, it has been speculated that gaps and rings are caused by young forming planets that have reached the pebble-isolation mass. Once they reach the pebble-isolation mass, these young forming planets perturb the disc and open a gap which generates a pressure bump outside of their orbits preventing the further inward flux of large pebbles. 

Many works have investigated the fate of these putative planets as they undergo gas-driven migration and accretion according to our current understanding of planet formation, and have analysed the synthetic planet population in comparison with the observed exoplanet sample. Past works have initially only considered the case of single-planets (i.e.\ no planet-planet dynamical interactions within the system, although multiple gaps are commonly observed, and multiple planet systems are not uncommon; \citep{lodato2019newborn,ndugu2019observed}), or multi-planet evolution for the specific case of the HD163296 system \citep{muller2022emerging} (a 1.5 solar mass star).

In this work we performed a more comprehensive N-body analysis where we simulated systems with 3, 4, 5 and 7 planets to study their growth, migration and also the gravitational interaction with the surrounding disc as well as the interactions between them. We located the planets at the positions where the gaps have been observed \citep{huang2018disk} and consider that our planets have just reached the pebble isolation mass. We investigated the effects of the viscosity parameter assuming three different values of $10^{-4}, 3.16 \cdot 10^{-4}$ and $10^{-3}$. In order to study the evolution of the planetary systems after the dissipation of the disc we continued to evolve the systems for 100Myr in total, while modeling the evolution of the disc for a few Myr.

We found that our simulations show an eccentricity distribution that is extremely low compared to the one observed for RV giants. We attribute this result to the low interaction between the planets which prevents scattering events and leads to low eccentricities. To be more precise, it is difficult to reproduce the observed eccentricity distribution by assuming 3 or 4 planets located at the positions of the gaps. 

Higher values of viscosity cause the planets to migrate more and accrete at higher rates resulting in a more massive and close-in population of giants. Although higher values of viscosity lead to higher final eccentricities, even the highest value of $\alpha=10^{-3}$ did not reproduce eccentricities large enough to match the observations. In addition, we did not find any strong dependence on the surface density profile. 
    
Finally, our results show that multi-ring discs cannot be explained by a one-planet-per-gap assumption. We found that systems consist of five and seven planets become instantly unstable. This indicates that the observed gaps and rings cannot be caused only by the presence of planets above pebble isolation mass. Additional processes (such as the ones mentioned in Sect.\ \ref{dis:unstable}), must be taken into account in order to understand the observed disc substructures. Thus, except in those cases where additional independent observations such as gas kinematics corroborate the presence of young accreting planets, one cannot attribute all disc substructures to planets in a bijective manner.

Future observations will be able to show if a yet-undetected long-period populations of planets actually exists, or if the putative planets located at the gaps in the observed young discs are not the progenitors of the current exoplanet sample.

\begin{acknowledgements}

B.B., thanks the European Research Council (ERC Starting Grant 757448-PAMDORA) for their financial support. \textbf{We thank the referee Ruobing Dong for his report that helped to clarify and improve our manuscript.}

\end{acknowledgements}

\bibliographystyle{aa} 
\bibliography{ref.bib}

\begin{appendix} \label{appendix}

\section{Distribution of mass and semi-major axis}

\begin{figure*}
	\centering
	\begin{minipage}{\columnwidth}
	   \centering
		\includegraphics[width=\hsize]{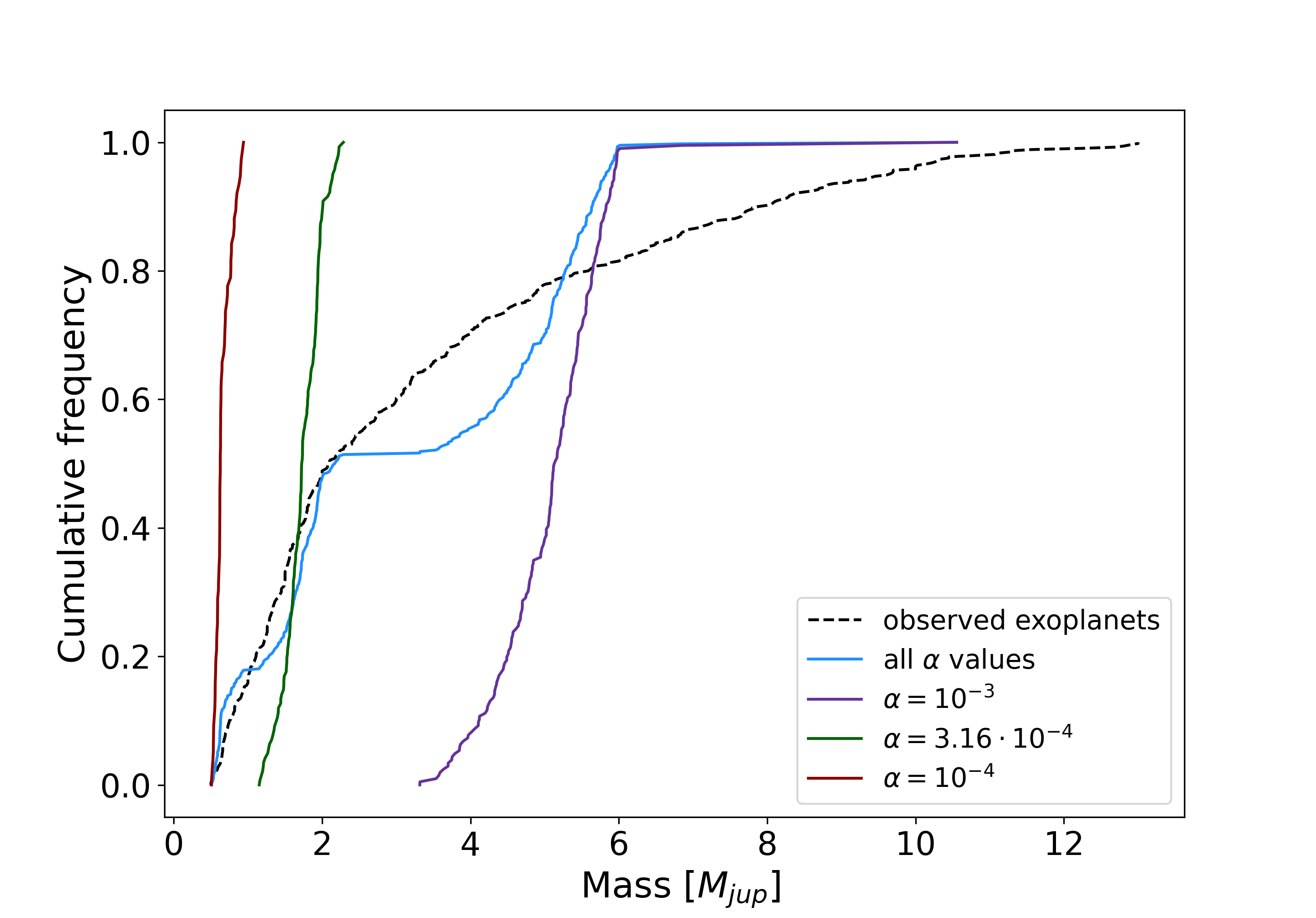}
	\end{minipage}%
	\begin{minipage}{\columnwidth}
		\centering
		\includegraphics[width=\hsize]{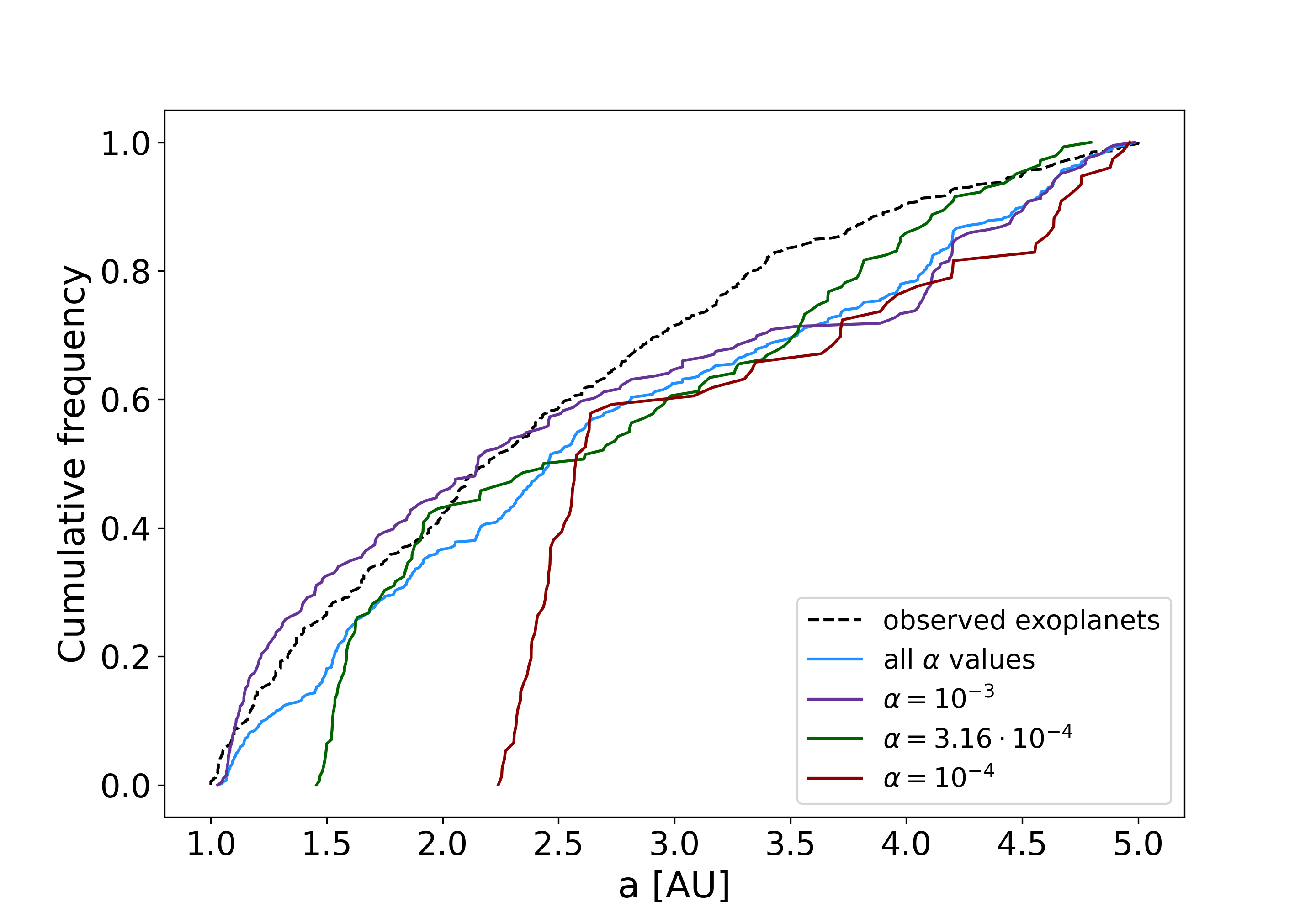}
    \end{minipage}
\caption{Cumulative distribution of mass (left panel) and semi-major axis (right panel) for the three-planet system. The plots include planets for which the semi-major axis is within the range of $1 \leq a[AU] \leq 5$ and mass between $0.5  \leq m[M_{Jup}] \leq 13 $.}
\label{fig:3pl_app}
\end{figure*}

\begin{figure*}
	\centering
	\begin{minipage}{\columnwidth}
		\centering
		\includegraphics[width=\hsize]{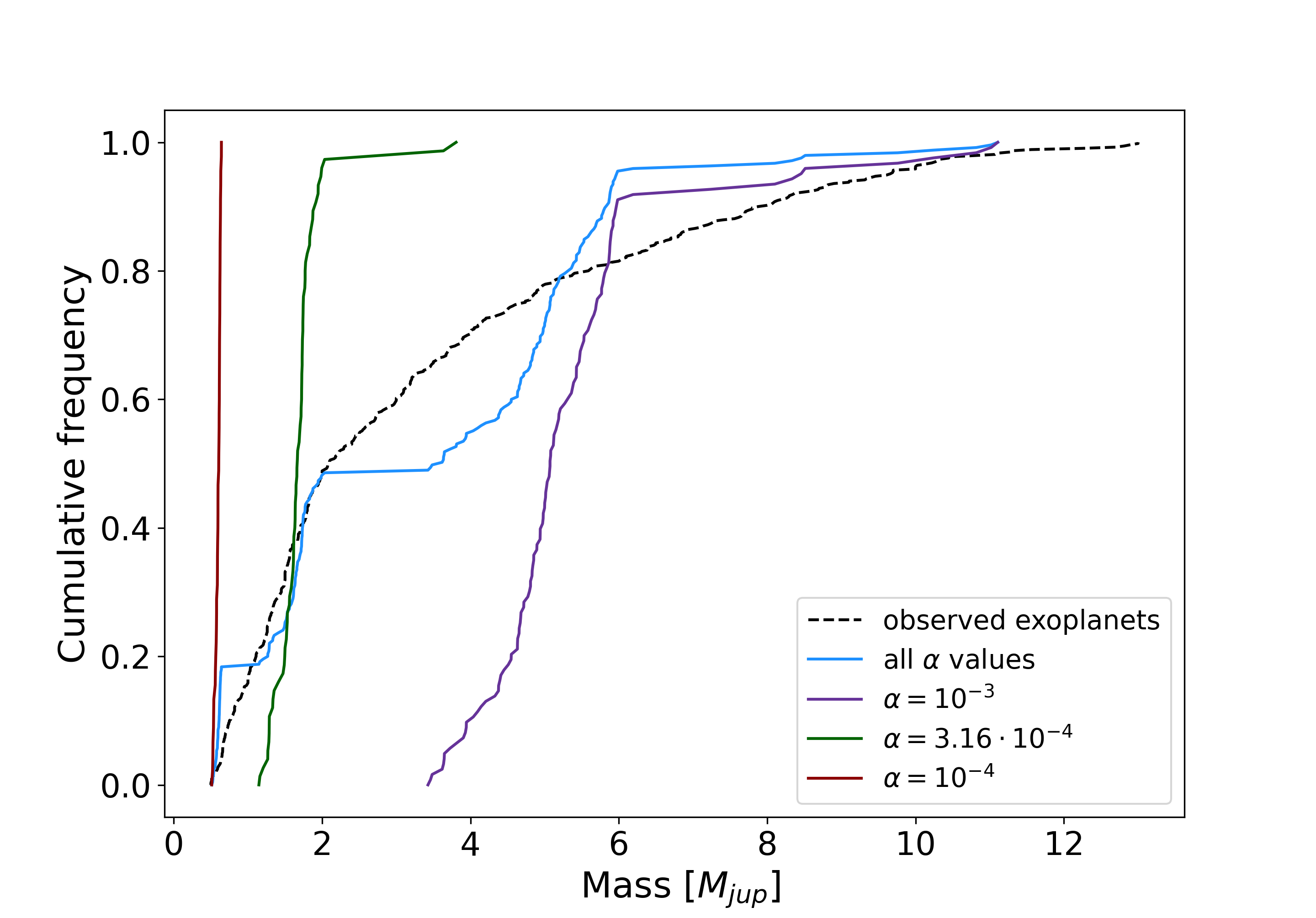}
	\end{minipage}%
	\begin{minipage}{\columnwidth}
		\centering
		\includegraphics[width=\hsize]{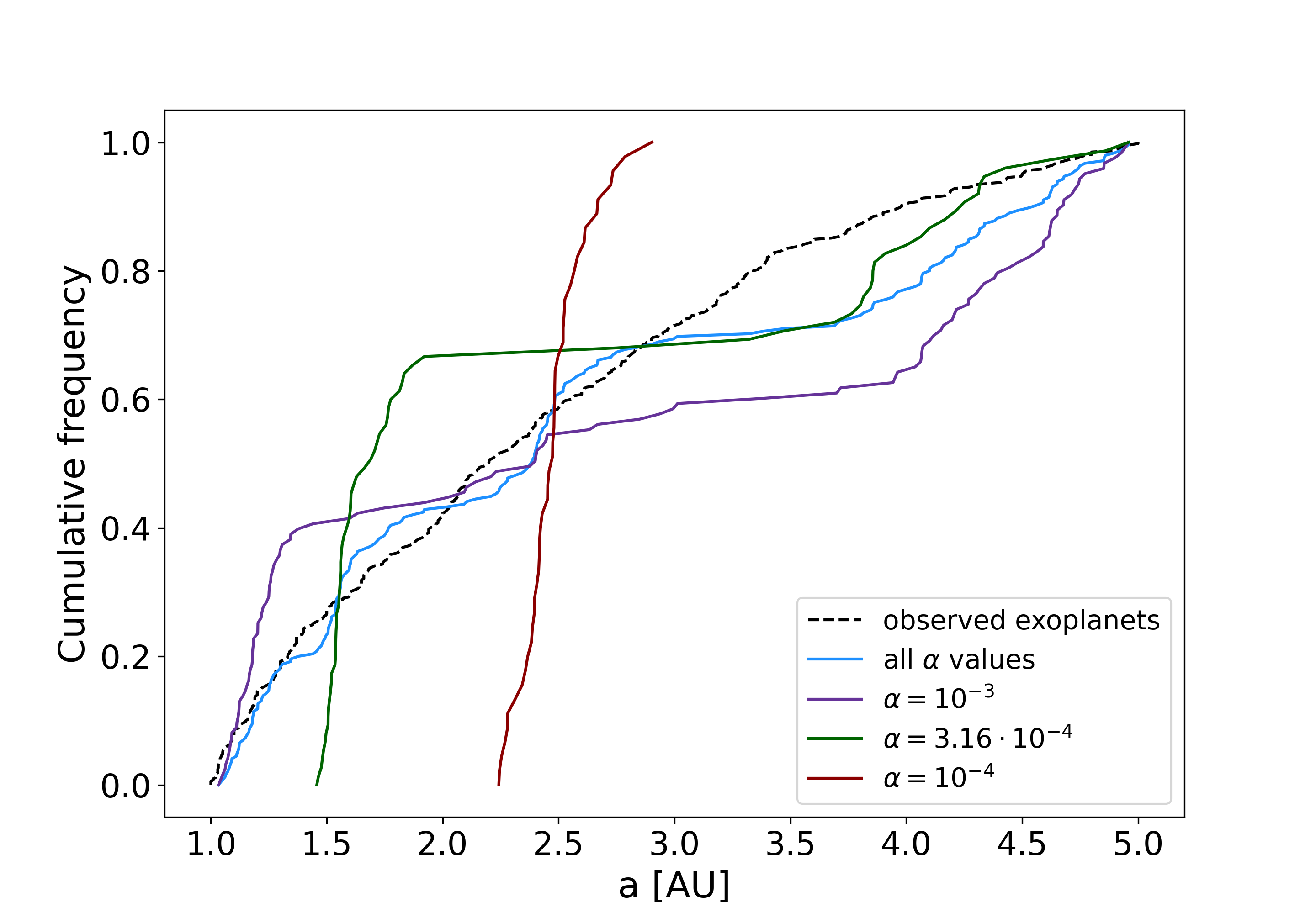}
  
    \end{minipage}
 \caption{Similar to Fig. \ref{fig:3pl_app}, but for systems with initially four planets.}   
 \label{fig:4pl_app}
\end{figure*}

We present here the cumulative distribution of mass and semi-major axis for systems with three (Fig. \ref{fig:3pl_app}) and four planets (Fig. \ref{fig:4pl_app}). We limited the dataset to include planets with final semi-major axis between [1-5]AU and final mass between [0.5-13]$M_{Jup}$, similar to the data set we presented in Sect. \ref{results} for the cumulative distribution of the eccentricity. The left panels show the cumulative distribution of the mass while the right panels show the cumulative distribution of the semi-major axis. We show the distribution for each value of the viscosity separately as well as for the combined dataset. For both systems we see that for higher values of viscosity the planets experience gas accretion and migration more efficiently. For the combined dataset (blue line) the final mean mass and mean semi-major axis is (3.16 $\pm$ 1.98)$M_{Jup}$ and (2.70 $\pm$ 1.18)AU for the three-planet system and (3.33 $\pm$ 2.31)$M_{Jup}$ and (2.56 $\pm$ 1.25)AU for the four-planet system. The mean mass of the observed giant planets is (3.35 $\pm$ 2.85)$M_{Jup}$ whereas the mean semi-major axis is (2.40 $\pm$ 1.07)AU. We see that both systems, with three and four planets, have similar mass and semi-major axis distribution as the observations, when taking into account the combined set. In general though, we must note that our systems cannot match the observations since our simulated systems did not reproduce the observed eccentricity distribution (see Sect. \ref{results} and Sect. \ref{discussion}).

\end{appendix}

\end{document}